\documentclass[nofootinbib]{revtex4}
\usepackage{graphicx}
\usepackage{latexsym}
\def\be{\begin{equation}}
\def\ee{\end{equation}}
\def\bea{\begin{eqnarray}}
\def\eea{\end{eqnarray}}

\begin{document}

\title{Note on the production of scale-invariant entropy perturbation in the Ekpyrotic universe}

\author{Mingzhe Li}
\email{limz@ustc.edu.cn}
\affiliation{
Interdisciplinary Center for Theoretical Study, University of Science and Technology of China, Hefei, Anhui 230026, China}


\begin{abstract}
In the standard entropic mechanism adopted in the simple Ekpyrotic models to generate the nearly scale-invariant and Gaussian primordial perturbation, the entropy direction is tachyonically unstable.
In this paper, we consider the stable production of the scale-invariant entropy perturbation in the Ekpyrotic universe via non-minimal couplings. In this model the non-minimally coupled massless scalar field serves as a spectator and
is stabilized by the introduced non-minimal couplings. It always corresponds to the entropy field during the contraction and with appropriate couplings can obtain a scale-invariant spectrum. This scenario requires
additional mechanisms such as curvaton or modulated preheating to convert the entropy perturbation to the curvature perturbation after the bounce.
\end{abstract}

\maketitle

\hskip 1.6cm PACS number(s): 98.80.Cq, 98.80.Bp \vskip 0.4cm

\hfill USTC-ICTS-13-07

\section{Introduction}

The Ekpyrotic and Cyclic models \cite{Ekpyrotic1,Ekpyrotic2,Ekpyrotic3}, as an alternative to
inflation \cite{Guth:1980zm,Linde:1981mu,Albrecht:1982wi}, provide not only solutions to the horizon and flatness problems existed in the
hot big bang cosmology but also mechanisms to generate initial perturbations for structure formation.
The recent results \cite{Ade:2013ktc,Ade:2013zuv,Ade:2013uln,Ade:2013ydc} from the Planck satellite confirmed further that the
primordial density perturbation is adiabatic, nearly scale-invariant and satisfies Gaussian statistics.
Apart from possible anomalies on very large angular scales (those anomalies have been observed by
WMAP satellite and confirmed by Planck) \cite{Ade:2013nlj}, the results could be well interpreted by the simple inflation models with single slow-rolling scalar field. A question one may ask is whether such a primordial perturbation can be produced in the Ekpyrotic/Cyclic universe in a smooth way.

In the Ekpyrotic model the universe is assumed to have experienced a slow contracting phase before bouncing to the hot expansion. An ultra slow contraction (Ekpyrotic phase) driven by ultra-stiff matter (the equation of state $w>1$) is needed to explain the smoothness and flatness of the universe and to suppress the BKL anisotropies \cite{BKL}. The productions of super-Hubble density perturbations are due to the fact that during the contraction the Hubble
radius was shrinking and the quantum vacuum fluctuations created deep inside it were able to cross the Hubble radius to outside regions.
It is well known that the original Ekpyrotic model \cite{Ekpyrotic1,Ekpyrotic2} with single scalar field cannot generate the required scale invariant adiabatic perturbation, the resulting spectrum for the curvature perturbation is strongly blue \cite{Ekpyrotic4,Ekpyrotic5,Ekpyrotic6,Ekpyrotic7} (see \cite{cai} for recent discussions of generation of scale-invariant curvature perturbation during the matter contraction preceding the Ekpyrotic phase).
Currently the best understood way to produce the right primordial perturbations is the so-called ``entropic mechanism" \cite{Ekpyroticentropy1,Lehners:2008qe} (for early references see \cite{Ekpyroticentropy2,Ekpyroticentropy3}) which was recently shown to be consistent with the Planck data \cite{Lehners:2013cka}.
With this mechanism multiple fields are introduced and it firstly generates entropy (or isocurvature) perturbations which might be scale-invariant. Then these entropy perturbations are
converted into curvature perturbations with identical spectra. The conversion is usually assumed to be finished before the bounce. 
However, for the standard entropic mechanism where all the scalar fields are canonical and minimally coupled to the gravity, the background evolution (scaling solution) is not stable, or the field which represents the entropy direction has a tachyon potential \cite{koyama1,koyama2,tolley}. This instability is needed
to keep the scale invariance of the entropy perturbation, and in some cases it is necessitated to convert the entropy perturbations to curvature ones at late time of the contracting phase. But this
instability also requires the tuning of the initial conditions or existence of
a preceding phase to keep the universe undergoing a long enough period of slow contraction so that the produced entropy perturbation has a
flat spectrum over extensive scales.

In this paper we consider the generation of scale-invariant entropy perturbation in a stable Ekpyrotic phase which admits scaling solutions.
Using the non-minimal couplings to the kinetic term of a massless scalar field to produce scale-invariant perturbations in contracting universe has been considered in Ref. \cite{nonminimal1,nonminimal2}.
In this paper, differing from the previous considerations, we focus on the cases in which the non-minimally coupled and massless scalar field serves as a spectator.
The non-minimal coupling brings a friction effect which damps the background velocity of the spectator field. If the friction is large enough the spectator field gets frozen quickly in the contracting universe. It always corresponds to the entropy direction and has no contribution to the evolution of the universe before the bounce. We will show that in some cases the spectator field lives in an effective de Sitter spacetime and obtains
a scale-invariant spectrum for its perturbations, similar to the pseudo Conformal universe \cite{conformal1,conformal2} and the Galileon Genesis \cite{galileon} (see also slow expansion scenario for different case \cite{piao}). Another similarity to the pseudo Conformal universe and the Galileon Genesis is that this model needs
extra mechanisms such as the curvaton \cite{curvaton} or modulated preheating \cite{modulation1,modulation2,modulation3} to convert the entropy perturbation to the curvature perturbation after the bounce (see also the application of curvaton mechanism to some non-singular bouncing universe \cite{qiu}). Furthermore, in our model the requirement of $w\gg 1$ which is necessary in the standard entropic mechanism can be relaxed.

This paper is organized as follows. In Sec. II, we discuss briefly the tachyonical instability presented in the standard entropic mechanism. In Sec. III, we discuss how to stabilize the entropy direction using the non-minimal
couplings, and with two models we show that the resulted spectra for the entropy perturbations are scale-invariant. Sec. IV is the conclusion.

\section{tachyonical instability of the standard entropic mechanism}

In the Ekpyrotic/Cyclic models, the universe experienced an ultra slow contraction with the equation of state $w>1$, during which the universe is smoothed and flattened and its background space-time is approximated the spatially flat Friedmann-Robertson-Walker (FRW) type which is conformal to the Minkowski space,
\be\label{FRW}
ds^2=a^2(\eta) \eta_{\mu\nu} dx^{\mu}dx^{\nu}~.
\ee
After that the universe enters into the contracting phase dominated by the kinetic terms of scalar fields, $w=1$, then bounces into the standard big bang expansion through the singularity.
In the simplest version of the standard entropic mechanism, the Ekpyrotic phase is governed by general relativity and two minimal coupled scalar fields with negative steep potentials
\be\label{system1}
\mathcal{L}=\frac{1}{2}\partial_{\mu}\phi_1\partial^{\mu}\phi_1+\frac{1}{2}\partial_{\mu}\phi_2\partial^{\mu}\phi_2+V_1
e^{-\frac{c_1}{M_p}\phi_1}+V_2e^{-\frac{c_2}{M_p}\phi_2}~,
\ee
where the parameters $V_i$ and $c_i$ for $i=1,~2$ are positive and $M_p=1/\sqrt{8\pi G}$ is the reduced Planck mass.
 This system has the scaling solutions for which the ratios $\phi_1'/\mathcal{H}$ and $\phi_2'/\mathcal{H}$ are constants, where the prime is the derivative with respect to the conformal time $\eta$ and
$\mathcal{H}=a'/a$ is the reduced Hubble rate which is negative in the contracting phase.
For such a double-field system, it is useful to project the linear perturbations of the scalar fields into the adiabatic direction $\sigma$ which parallels the trajectory of the background fields' motion and the orthogonal,
entropy direction $s$ \cite{Gordon:2000hv}. The adiabatic field $\sigma$ is defined so that in the background
\be
\sigma'^2=\phi_1'^2+\phi_2'^2~,
\ee
or
\be
\sigma'=\phi_1'\cos\theta+\phi_2'\sin\theta
\ee
with $\cos\theta=\phi_1'/\sigma',~\sin\theta=\phi_2'/\sigma'$. Because the entropy direction is perpendicular to the trajectory, its background does not evolve. This can be seen from the
following equation which guarantees the orthogonality of $s$ and $\sigma$ at any time
\be
s'=-\phi_1'\sin\theta+\phi_2'\cos\theta=\frac{-\phi_1'\phi_2'+\phi_2'\phi_1'}{\sigma'}=0~.
\ee
The background part of the entropy field is stabilized but its perturbation can evolve.
The projection of the perturbations along these two orthogonal directions are similarly
\bea
&& \delta\sigma=\delta\phi_1\cos\theta+\delta\phi_2\sin\theta~,\nonumber\\
&&\delta s=-\delta\phi_1 \sin\theta+\delta\phi_2\cos\theta~.
\eea
The perturbation of the adiabatic field $\delta\sigma$ relates to the comoving curvature perturbation via $\mathcal{R}=\psi+\frac{\mathcal{H}}{\sigma'}\delta\sigma$ which seeds the formation of the large scale structure of the universe, where
$\psi$ is the curvature perturbation to the constant time three-surface\footnote{More precisely, it is the curvature perturbation of the uniform density surface $-\zeta=\psi+\frac{\mathcal{H}}{\rho'}\delta\rho$ which takes the central roll in initiating the structure formation. On super Hubble scales, $\mathcal{R}$ and $-\zeta$ are identical.}.
The entropy perturbation $\delta s$ decouples from the metric perturbations at large scales, but if the background trajectory is curved, $\theta'\neq 0$, it provides a source to the adiabatic perturbation \cite{Gordon:2000hv}.

Generally these definitions only have instantaneous meanings because the angle $\theta$ is not
constant and it depends on the background fields' evolution.
If the system dwells in the scaling solutions, the background fields' motion is along a straight line in the field space, $\theta$ does not change with time, the adiabatic and entropy fields can be considered as fields' redefinitions by
a rotation in the field space \cite{Ekpyroticentropy3}
\bea
& &\sigma=\phi_1\cos\theta+\phi_2\sin\theta\nonumber\\
& &s=-\phi_1\sin\theta+\phi_2\cos\theta~,
\eea
where $\sigma$ and $s$ contain the background and perturbations.
But we should keep in mind that $s$ should be stabilized if it represents the true entropy direction.
The system (\ref{system1}) only possesses one scaling solution for which the generated entropy perturbation is scale-invariant. In this
scaling solution \cite{Ekpyroticentropy1,koyama1},
\be\label{scaling}
\frac{\phi_1'}{\phi_2'}={c_2  \over c_1}~,~w=\frac{c_1^2c_2^2}{3(c_1^2+c_2^2)}-1
\ee
and
\be
\sin\theta=\frac{c_1}{\sqrt{c_1^2+c_2^2}}~,~\cos\theta=\frac{c_2}{\sqrt{c_1^2+c_2^2}}~.
\ee
With these equations the action for $s$ can be reformulated as
\be
S_s=\int d^4x\sqrt{g} [{1\over 2}g^{\mu\nu}\partial_{\mu}s\partial_{
\nu}s-V(\sigma,~s)]
\ee
with
\be
V(\sigma,~s)=-\exp(-\frac{c_1c_2}{\sqrt{c_1^2+c_2^2}M_p}\sigma) [V_1\exp(\frac{c_1^2}{\sqrt{c_1^2+c_2^2}M_p}s)+V_2\exp(-\frac{c_2^2}{\sqrt{c_1^2+c_2^2}M_p}s)]~.
\ee
Because $V_i$ and $c_i$ for $i=1,~2$ are positive, the potential $V(\sigma,~s)$ is negative with upper bound. We can only have $s'=0$ at the maximum of the potential along $s$ direction.
That is to say, the entropy field has a tachyonic potential and $s$ cannot be stabilized at its top. This also
indicates that the scaling solution (\ref{scaling}) is unstable. If $s$ is away from the maximum of its potential,
it will not represent the instantaneous entropy direction. In order to keep the background of $s$ at constant for enough
long time, one must tune the initial conditions or assume a preceding phase to take the system to the position very close to the scaling solution (\ref{scaling}).

We will see that the tachyonic instability is necessary for this system to generate a scale-invariant
entropy perturbation. Assuming the system dwells in the (unstable) scaling solution, the entropy field $s$ stays at rest at the top of the potential $V(\sigma,~s)$ along $s$-direction.
Its fluctuation has the quadratic action\footnote{Usually we should distinguish the background of the entropy field and its fluctuation so that $s=s_0+\delta s$. Because $s_0$ is a constant and
we can shift it to zero, so $s$ only refers to fluctuation.}
\be
S_s={1\over 2}\int d^4x a^2[\eta^{\mu\nu}\partial_{\mu}s\partial_{\nu}s -a^2m_s^2 s^2]~,
\ee
where $\eta^{\mu\nu}$ is the Minkowski metric and $m_s$ is the effective mass, $m_s^2=\frac{\partial^2 V(\sigma,~s)}{\partial s^2}$.
Redefining $v=a s$, it satisfies the equation (in Fourier space):
\be\label{s1}
v''+(k^2-\frac{a''}{a}+a^2m_s^2)v=0~.
\ee
For the universe with constant equation of state $w$ (scaling solution), it is easy to obtain
\be
\frac{a''}{a}=\frac{1-3w}{(1+3w)^2}\frac{2}{\eta^2}
\ee
from the Friedmann equation $\mathcal{H}^2=\frac{a^2}{3M_p^2}\rho$, where $\rho$ is the energy density of the universe and the conformal time is negative $-\infty <\eta<0$ if we set the bouncing time at $\eta=0$.
Scale-invariant spectrum of $s$ requires the terms in Eq. (\ref{s1}) has the form
\be
\frac{a''}{a}-a^2m_s^2=\frac{2}{\eta^2}~,
\ee
hence
\be
a^2m_s^2=[\frac{1-3w}{(1+3w)^2}-1]\frac{2}{\eta^2}~.
\ee
In the Ekpyrotic phase $w>1$, this means that $m_s^2<0$, the entropy field $s$ has a tachyonic mass.
Detailed studies \cite{Ekpyroticentropy1} have shown that for the system (\ref{system1}) the scale-invariant entropy perturbation can be generated
if $w\gg 1$. For this system in some considerations another requirement of the tachyonic stability is that it provides a way
to bend the trajectory of the background field's motion and convert the produced entropy perturbation to the curvature perturbation.

\section{The stabilization of the entropy field with non-minimal couplings}

In this section we will consider the non-minimal couplings in the double-field system and apply it to the Ekpyrotic universe. In our consideration the first scalar field is still canonical and has
a negative exponential potential but the second field is massless and its kinetic term non-minimally coupled to the first one.
The full Lagrangian is
\be\label{lagrangian2}
\mathcal{L}={1\over 2}\partial_{\mu}\phi\partial^{\mu}\phi+V_0e^{-\frac{\lambda}{M_p}\phi}+{1\over 2}e^{-\frac{\alpha}{M_p}\phi}\partial_{\mu}\chi\partial^{\mu}\chi~.
\ee
Such a model was considered by the authors of Ref. \cite{nonminimal1} in the Pre-Big-Bang cosmology \cite{pre1,pre2} and $\chi$ was treated as an axion-like field.
The perturbations of this system were studied in more general expanding or contracting background in Ref. \cite{nonminimal2}. The non-linear perturbations of the two-field models with non-minimal couplings were
studied in \cite{RenauxPetel:2008gi}. 
In this paper we will focus on the case where $\alpha=\lambda$ so that
\be\label{lagrangian}
\mathcal{L}={1\over 2}\partial_{\mu}\phi\partial^{\mu}\phi+e^{-\frac{\lambda}{M_p}\phi}({1\over 2}\partial_{\mu}\chi\partial^{\mu}\chi +V_0)~,
\ee
with $V_0>0$ and $\lambda>0$. This model shows a dilaton type coupling to the system of a massless scalar field plus a negative cosmological constant.
The Euler-Lagrange equations for these two coupled fields have the form
\bea
& &\Box \phi+\frac{\lambda}{M_p}e^{-\frac{\lambda}{M_p}\phi}({1\over 2}\partial_{\mu}\chi\partial^{\mu}\chi +V_0)=0~,\\
& &\Box \chi-\frac{\lambda}{M_p}\partial_{\mu}\phi\partial^{\mu}\chi=0~,\label{equationchi}
\eea
where $\Box$ is the D'Alembert operator.
The above equations of motion in the FRW universe (\ref{FRW}) together with the Friedmann equation
\be
\mathcal{H}^2=\frac{1}{6M_p^2}(\phi'^2+\chi'^2e^{-\frac{\lambda}{M_p}\phi}-2a^2V_0e^{-\frac{\lambda}{M_p}\phi})
\ee
can be reformulated as
\bea\label{equation}
& &\dot x=3x(x^2+y^2-1)-\frac{\sqrt{6}}{2}\lambda (x^2+2y^2-1)\nonumber\\
& &\dot y=3y(x^2+y^2-1)+ \frac{\sqrt{6}}{2}\lambda xy~,
\eea
where dimensionless variables $x,~y$ are defined as  \cite{nonminimal2}
\be
x\equiv \frac{\phi'}{\sqrt{6}M_p\mathcal{H}}~,~y\equiv \frac{e^{-\frac{\lambda}{2M_p}\phi}\chi'}{\sqrt{6}M_p\mathcal{H}}~,
\ee
and $\dot x\equiv dx/d\ln a,~\dot y\equiv dy/d\ln a$. We can also define another dimensionless parameter
\be
z\equiv \frac{a\sqrt{V_0}e^{-\frac{\lambda}{2M_p}\phi}}
{\sqrt{3}M_p\mathcal{H}}~,
\ee
but it is constrained by the Friedmann equation $x^2+y^2-z^2=1$, so $x,~y$ are the independent phase space variables except the constraint $x^2+y^2\geq 1$.

Eqs. (\ref{equation}) have three fixed points, i.e., $(x_0=-1,~y_0=0)$, $(x_0=1,~y_0=0)$ and $(x_0=\lambda/\sqrt{6}, y_0=0)$. These fixed points correspond to scaling solutions.
Whether these solutions are stable depends on the behaviors of the small deviations away from them. The small deviations $X=x-x_0$ and $Y=y-y_0$ satisfy the linear equations
\bea
\dot X=(9x_0^2-\sqrt{6}\lambda x_0-3) X~,~\dot Y=(3x_0^2+\frac{\sqrt{6}}{2}\lambda x_0-3) Y~,
\eea
and from these equations one obtains that
\be
|X|\sim \exp[(9x_0^2-\sqrt{6}\lambda x_0-3)\ln a]~,~|Y|\sim \exp[(3x_0^2+\frac{\sqrt{6}}{2}\lambda x_0-3)\ln a]~.
\ee
Because in contracting universe $\ln a$ is decreasing, stable solutions require both coefficients in the exponentials
$9x_0^2-\sqrt{6}\lambda x_0-3$ and $3x_0^2+\frac{\sqrt{6}}{2}\lambda x_0-3$ be positive.
With the assumption that $\lambda>0$, we can see that the first fixed point $(x_0=-1,~y_0=0)$ is always unstable. The second fixed point
$(x_0=1,~y_0=0)$ corresponds to an attractor solution (stable) for $\lambda \leq \sqrt{6}$ and the third fixed point also corresponds to an attractor solution in the case $\lambda >\sqrt{6}$.

From now on we only consider these two attractor solutions. 

(1) $x_0=1,~y_0=0,~0<\lambda\leq \sqrt{6}$, at this point $z_0=0$, the potential term and the kinetic energy of $\chi$ are negligibly small. The universe is dominated by the kinetic term of $\phi$. The equation of state is $w=1$. This corresponds to $\phi\rightarrow +\infty$ and made the field $\chi$ non-dynamical. Such a phase can only marginally suppress the BKL anisotropies. Like the single field Pre-Big-Bang universe, there is no entropy perturbation, but the produced adiabatic perturbation has a strongly blue tilt spectrum. 

(2) $x_0=\frac{\lambda}{\sqrt{6}}>1, ~y_0=0$, at this point $z_0=-\sqrt{\frac{\lambda^2}{6}-1}$. Both the kinetic and potential energies of
$\phi$ scales as $a^{-3(1+w)}$. The equation of state of the universe is $w=\frac{\lambda^2}{3}-1>1$. The field $\chi$ is stabilized with $\chi'=0$, this is due to the damping effect brought by the exponential coupling in the equation of motion (\ref{equationchi}),
\be
\chi''+2\mathcal{H}\chi'-\frac{\lambda}{M_p}\phi'\chi'=\chi''-(\sqrt{6}\lambda x_0-2)\mathcal{H}\chi'
=\chi''-(\lambda^2-2)\mathcal{H}\chi'=0~.
\ee
Without the non-minimal coupling $\chi'$ increases as $a^{-3}$. When non-minimal coupling is introduced, it brings an extra friction term
and because $\lambda^2>6$, the velocity $\chi'$ decreases and quickly approaches zero. In the non-minimal coupling case, the definitions of adiabatic and entropy directions have the generalized form \cite{nonminimal2}
\be
\sigma'^2=\phi'^2+\chi'^2e^{-\frac{\lambda}{M_p}\phi}~,
\ee
and 
\be
\sigma'=\phi'\cos\theta+\chi'e^{-\frac{\lambda}{2M_p}\phi}\sin\theta~,
\ee
with 
\be
\cos\theta=\frac{\phi'}{\sigma'}~,~\sin\theta=\frac{\chi'e^{-\frac{\lambda}{2M_p}\phi}}{\sigma'}~.
\ee
The stabilization of the field $\chi$ leads to $\theta=0$, and $\chi$ always represents the entropy direction. The adiabatic and entropy perturbations are decoupled, the conversion from the entropy perturbation to the adiabatic one can only happen at later time. These can be seen clearly from the linear perturbation equation of $\chi$ field.  
The linear perturbation equation of the fluctuation $\delta\chi=\chi(\eta,~\vec{x})-\chi(\eta)$ around the frozen background $\chi'(\eta)=0$ can be obtained from the quadratic action for $\delta\chi$. When all the perturbations are included
the action for the $\chi$ field is 
\be
S_{\chi}={1\over 2}\int d^4x (\sqrt{g}+\delta\sqrt{g}) e^{-\frac{\lambda}{M_p}(\phi+\delta\phi)}(g^{\mu\nu}+\delta g^{\mu\nu})\partial_{\mu}(\chi+\delta\chi)\partial_{\nu}(\chi+\delta\chi)~,
\ee
and expanding it to the second order by considering $\partial_{\mu}\chi=0$, we can read out the quadratic action for $\delta\chi$ 
\be\label{effectiveaction1}
S_{\delta\chi}={1\over 2}\int d^4x \sqrt{g} e^{-\frac{\lambda}{M_p}\phi}g^{\mu\nu}\partial_{\mu}\delta\chi\partial_{\nu}\delta\chi~.
\ee
Because the exponential factor which couples to the kinetic term of $\chi$ is the same as the potential of $\phi$,
and from
\be
z_0^2= \frac{a^2 V_0e^{-\frac{\lambda}{M_p}\phi}}
{3M_p^2\mathcal{H}^2}=\frac{\lambda^2}{6}-1~,~e^{-\frac{\lambda}{M_p}\phi}=\frac{(\lambda^2-6)M_p^2\mathcal{H}^2}{2a^2V_0}~,
\ee
 the action (\ref{effectiveaction1}) is
\bea\label{effectiveaction}
S_{\delta\chi}&=&{1\over 2}\int d^4x a^2e^{-\frac{\lambda}{M_p}\phi}\eta^{\mu\nu}\partial_{\mu}\delta\chi\partial_{\nu}\delta\chi\nonumber\\
&=&{1\over 2}\int d^4x \frac{(\lambda^2-6)M_p^2\mathcal{H}^2}{2V_0}\eta^{\mu\nu}\partial_{\mu}\delta\chi\partial_{\nu}\delta\chi~,
\eea
where $\eta^{\mu\nu}$ is the Minkowski metric. With the constant equation of state $w=\frac{\lambda^2}{3}-1$, from the Friedmann equation
we can solve that
\be
\mathcal{H}=\frac{2}{1+3w}\frac{1}{\eta}=\frac{2}{\lambda^2-2}\frac{1}{\eta}~.
\ee
Substituting it into Eq. (\ref{effectiveaction}), we get
\be\label{effectiveaction2}
S_{\delta\chi}={1\over 2}\int d^4x \frac{1}{h^2\eta^2}\eta^{\mu\nu}\partial_{\mu}\delta\chi\partial_{\nu}\delta\chi
\ee
with
\be
h=\frac{\lambda^2-2}{M_p}\sqrt{\frac{V_0}{2(\lambda^2-6)}}~.
\ee
From these equations we can see that $\delta\chi$ is a massless field living in an effective de Sitter space with the metric
\be
\bar{g}_{\mu\nu}=\frac{1}{h^2\eta^2}\eta_{\mu\nu}~.
\ee
Similar to the cases of the pseudo Conformal universe \cite{conformal1,conformal2} and the Galileon Genesis \cite{galileon}, the field $\delta\chi$ will obtain a 
scale-invariant spectrum. This can be seen more clearly from the following discussions. 
By field redefinition $u=\delta\chi/(h\eta)$ and integration by parts the quadratic action (\ref{effectiveaction2}) may be reformulated as
\be
S_{\delta\chi}={1\over 2}\int d^4x [\eta^{\mu\nu}\partial_{\mu}u\partial_{\nu}u+\frac{2}{\eta^2}u^2]~.
\ee
The field $u$ lives in the Minkowski space and has a time varying mass $m_u^2=-2/\eta^2$. Its equation of motion for fixed mode is 
\be
u_k''+(k^2-\frac{2}{\eta^2})u_k=0~.
\ee
This equation has two independent solutions, i.e., the solution has the general form
\be\label{modefunction}
u_k=C_1(k) e^{-ik\eta}(1-\frac{i}{k\eta})+C_2(k) e^{ik\eta}(1+\frac{i}{k\eta})~,
\ee
where $C_i(k)$ for $i=1,~2$ are integral constants which are functions of $k$ but independent of time. Usually we assume the quantum fluctuations are created from the 
Bunch-Davies vacuum at early time when $k|\eta|\gg 1$, for which the mode function has the form 
\be
{\rm lim}_{\eta\rightarrow -\infty} u_k=\frac{e^{-ik\eta}}{\sqrt{2k}}~.
\ee
With this initial vacuum selection the coefficients in Eq. (\ref{modefunction}) are fixed to $C_1(k)=1/\sqrt{2k}$, $C_2(k)=0$ and the mode function is $u_k=\frac{e^{-ik\eta}}{\sqrt{2k}}(1-\frac{i}{k\eta})$
which asymptotes to $\frac{-ie^{-ik\eta}}{\sqrt{2k^3}\eta}$ at late time when $k|\eta|\ll 1$. Hence the produced power spectrum of $\delta\chi$ is (remember $\delta\chi=h\eta u$)
\be
\mathcal{P}_{\delta\chi}^{1/2}\equiv \frac{k^{3/2}}{\sqrt{2}\pi}h\eta |u_k|=\frac{h}{2\pi}~,
\ee
which is scale-invariant and does not evolve with time outside the Hubble radius. 

One of the advantages of such models is that the condition $w\gg 1$ which is needed in the standard entropic mechanism is not necessary any more. What we need is a constant $w$ and it should be larger than $1$ to suppress the BKL anisotropies. We should point out that when the universe enters into the kinetic
term dominated contraction after the Ekpyrotic phase, the field $\chi$ is still stabilized and the trajectory of the background's motion is not curved during the whole contracting phase. So we need other mechanisms such as the curvaton or the modulated preheating to convert the scale-invariant entropy perturbation to the curvature one. 

Up to now we showed that the model (\ref{lagrangian}) can successfully generate exact scale-invariant perturbation for the entropy field during the Ekpyrotic phase. This model corresponds to the special case of the more general non-minimal coupling models (\ref{lagrangian2}). Let us comment on the more general model (\ref{lagrangian2}) in which $\alpha$ is different from $\lambda$, this seems to be more natural from the viewpoint of fundamental physics. As implied in Ref. \cite{nonminimal2}, the fixed point which corresponds to the stable Ekpyrotic contraction is the same with that discussed above for the special case, i.e., $x_0=\frac{\lambda}{\sqrt{6}}>1, ~y_0=0$ and $z_0=-\sqrt{\frac{\lambda^2}{6}-1}$. In the same way as before, the field $\chi$ is stabilized and always represents the entropy direction. However, the quadratic action for its fluctuation is different. A straightforward calculation shows that 
\be
S_{\delta\chi}={1\over 2}\int d^4x a^2e^{-\frac{\alpha}{M_p}\phi}\eta^{\mu\nu}\partial_{\mu}\delta\chi\partial_{\nu}\delta\chi
={1\over 2}\int d^4x q^2 \eta^{\mu\nu}\partial_{\mu}\delta\chi\partial_{\nu}\delta\chi~,
\ee
where $q\propto \eta^{\beta}$ with $\beta=-(\lambda \alpha-2)/(\lambda^2-2)$. Defining $u=q\delta\chi$, it is easy to find that $u$ satisfies the equation
\be
u_k''+(k^2-\frac{q''}{q})u_k=u_k''+(k^2-\frac{\beta (\beta-1)}{\eta^2})u_k=0~.
\ee
With the selection of the Bunch-Davies vacuum at early time when $k|\eta|\gg 1$, the solution to the above equation is 
\be
u_k=\sqrt{-\frac{\pi}{2}\eta} H_{\nu}^{(1)}(-k\eta)~,
\ee
where $H_{\nu}^{(1)}$ is the first kind Hankel function and $\nu=\sqrt{1/4+\beta(\beta-1)}$.  At late time when the perturbation mode is outside the Hubble radius, i.e., $k|\eta|\ll 1$, the mode function asymptotes 
$u_k\sim -\eta^{1/2} (-k\eta)^{-\nu}$. So that the power spectrum for $\delta\chi$ is 
\be
\mathcal{P}_{\delta\chi}\sim k^3 |\frac{u_k}{q}|^2\sim \frac{(-k\eta)^{3-2\nu}}{\eta^{2+2\beta}}~.
\ee
The spectral index is $n_s=4-2\nu$. If $\alpha$ only differs from $\lambda$ slightly, the number $p=\alpha-\lambda$ is small, one can obtain that 
\be
n_s\simeq 1-\frac{2\lambda p}{\lambda^2-2}~.
\ee
If $\alpha$ is larger than $\lambda$ slightly, $p$ is a small positive number, the produced spectrum has a small red tilt.

As a second example of the stable generation of the entropy perturbation during Ekpyrotic phase with scaling solutions we may consider the massless field $\chi$ non-minimally couples to the curvature scalar
\be\label{action}
S_{\chi}=-{1\over 2}\int d^4x \sqrt{g} \frac{R}{M^2}g^{\mu\nu}\partial_{\mu}\chi\partial_{\nu}\chi~,
\ee
where $M$ represents a cut-off mass and the minus sign at the right hand side is to keep the field away from the ghost. Similar to the previous case, this coupling can stabilize the field $\chi$ and make it a spectator so that it always corresponds to the entropy direction. 
The curvature scalar $R=6a''/a^3=6(\mathcal{H}'+\mathcal{H}^2)/a^2$, and in the Ekpyrotic universe where the total equation of state $w$ is a constant larger than $1$, with the equation
\be
\mathcal{H}=\frac{2}{1+3w}\frac{1}{\eta}~,
\ee
we get
\be
R=\frac{12(1-3w)}{(1+3w)^2a^2\eta^2}~.
\ee
Substituting it into the equation (\ref{action}),
we have
\be
S_{\chi}={1\over 2}\int d^4x \frac{1}{\tilde{h}^2\eta^2}\eta^{\mu\nu}\partial_{\mu}\chi\partial_{\nu}\chi~,
\ee
with
\be
\tilde{h}=\frac{(1+3w)M}{2\sqrt{3(3w-1)}}~.
\ee
Hence in this case the entropy field $\chi$ obtains a scale-invariant spectrum $\mathcal{P}_{\chi}^{1/2}=\tilde{h}/2\pi$. 
Same as before, it also needs extra conversion mechanisms after the bounce. If the universe directly bounces into the radiation dominated expanding phase in which $R$ vanishes and the field $\chi$ is not dynamical, it is better to consider the modulated preheating as the conversion mechanism.

\section{Conclusions}

In this paper, we considered the generation of entropy perturbation in the Ekpyrotic universe without tachyonical instability. In our model we introduced the non-minimally coupled massless scalar field. Its background is stabilized due to the damping of the non-minimal coupling. It serves as a spectator field and always corresponds to the entropy direction. With appropriate non-minimal couplings, this entropy field lives in an effective de Sitter space and obtains a scale-invariant spectrum for its perturbation. Because the curvature perturbation produced during the contraction with constant equation of state is strongly blue, its amplitude at large scales could be negligible and we can think that the curvature perturbation which seeds the structure formation at these scales totally comes from the conversion from the entropy perturbation. The model proposed in this paper has the advantage that we do not need the equation of state $w\gg 1$. The price we will take is that extra mechanisms are necessary to convert the scale-invariant entropy perturbation to the curvature one after the bounce. The efficiency of the conversion, however, is model dependent.

\section{Acknowledgement}

The author thanks Yun-Song Piao for useful discussions. This work is supported in part by National Science Foundation of China under Grants Nos. 11075074 and 11065004 and by SRF for ROCS, SEM.

{}

\end{document}